\documentclass[aps,prl,twocolumn]{revtex4}
\usepackage{amsbsy,latexsym,amsmath}
\usepackage{amsfonts}
\usepackage{amssymb}
\usepackage[mathscr]{eucal}
\usepackage{epsfig,graphics,graphicx}
\usepackage[colorlinks=true,urlcolor=blue,citecolor=blue]{hyperref}
\renewcommand{\vec}[1]{\mbox{\boldmath$#1$}}

\newcommand{\tr}{{\rm tr}\,}
\newcommand{\ket}[1]{\left|{#1}\right\rangle}
\newcommand{\bra}[1]{\left\langle{#1}\right|}
\newcommand{\braket}[2]{\langle{#1}|{#2}\rangle}
\newcommand{\ketbrad}[1]{\left|{#1}\rangle\!\langle{#1}\right|}
\newcommand{\ketbra}[2]{\left|{#1}\rangle\!\langle{#2}\right|}

\begin{document}

\title{Exact Identification of a Quantum Change Point}
\author{Gael Sent\'{\i}s$^{1,2}$}\email{gael.sentis@uni-siegen.de}
\author{John Calsamiglia$^{3}$}\email{John.Calsamiglia@uab.cat}
\author{Ramon Mu\~{n}oz-Tapia$^{3}$}\email{Ramon.Munoz@uab.cat}
\affiliation{
$^{1}$Naturwissenschaftlich-Technische Fakult\"at, Universit\"at Siegen, 57068 Siegen, Germany\\
$^{2}$Departamento de F\'{i}sica Te\'{o}rica e Historia de la Ciencia, Universidad del Pa\'{i}s Vasco UPV/EHU, E-48080 Bilbao, Spain\\
$^{3}$F\'{i}sica Te\`{o}rica: Informaci\'{o} i Fen\`{o}mens Qu\`antics, Departament de F\'{\i}sica, Universitat Aut\`{o}noma de Barcelona, 08193 Bellaterra (Barcelona), Spain}

\begin{abstract}
%
%
The detection of change points is a pivotal task in statistical analysis. 
In the quantum realm, it is a new primitive where one aims at identifying the point where a  source that supposedly prepares a sequence of particles in identical quantum states starts preparing a mutated one. We obtain the optimal procedure to identify the change point with certainty---naturally at the price of having a certain probability of getting an inconclusive answer.
We obtain the analytical form of the optimal probability of successful identification  for any length of the particle sequence. 
We show that the conditional success probabilities of identifying each possible change point show an unexpected oscillatory behaviour. 
We also discuss local (online) protocols and compare them with the optimal procedure.

\end{abstract}
\maketitle
%
%
We are surrounded by changes. In many  physical settings  there is a point when things start to be different from what they used to be.  This may be due to a permanent alteration that occurred at some previous time. 
The specific time when this mutation happened can have many significant practical consequences that make its identification crucial. In statistical analysis this problem, the detection of sudden changes in the characteristics of an observed process, 
is known as the change point problem, a vast field of research with many applications~\cite{page, brodsky,baseville}. 

The first extension 
of this problem into a quantum setting was introduced very recently in Ref.~\cite{prl-us}:  a source  that is supposed to prepare a sequence of quantum particles in some default state suffers an alteration at some unspecified point, 
after which it starts preparing a different, mutated state. 
Then, given a sequence of particles, the task is to detect when this change point has taken place.
In this bare-bones setting the initial and final states are assumed to be pure 
and known
and no prior information is given about the location of the change, 
that is, for a given sequence of length $n$,
every point in the sequence is equally likely to be the change point.
Remarkably, an analytical expression for the success probability of correct identification in terms of complete elliptic functions has been obtained in the asymptotic limit of 
long sequences.
Furthermore, it is shown 
to be a finite quantity that depends only on the overlap of the initial and final state.

The protocols devised 
in Ref.~\cite{prl-us} allow for errors,
that is, 
with some nonzero probability the
change point will be misidentified
and the optimal protocol is defined as the one that minimizes the rate of errors. 
There are, however,
situations where 
giving an erroneous answer is  inadmissible. In these,
one would take action only if the event is detected with absolute certainty, and otherwise remain idle.
The optimization then consists in maximizing the rate of correct identification under the constraint that no errors are made, or, equivalently,
in minimizing the rate of inconclusive outcomes, i.e., those that do not provide a certain answer \cite{unambiguous}. 
In the context of quantum state discrimination the first approach is known as minimum-error discrimination, whereas the second is termed unambiguous discrimination.
Very much like it happens in the minimum-error approach, there are very few examples 
of unambiguous discrimination scenarios with a complete analytical solution:
beyond two hypotheses they 
reduce essentially to very symmetric cases~\cite{chefles} (see also Refs.~\cite{sband,pangwu}). 
Exceptionally, the unambiguous detection of quantum change points is one of the unique cases involving multiple hypotheses that can be solved completely. 
In this Letter, we find the optimal measurement and the optimal success probability for unambiguous detection of quantum change points for any possible pair of default and mutated states and sequences of arbitrary length.

The path to finding the analytical solution relies on 
formulating the problem as a semidefinite program (SDP)~\cite{sdp}.
SDPs are a type of efficiently solvable convex optimization problems that admit linear constraints over matrix variables.
Each SDP has a primal and a dual version whose feasible sets provide lower and upper bounds on the solution of the optimization, certifying its level of accuracy.
Generic formulations of state discrimination tasks as SDPs can be found in the literature \cite{nakahira,eldar,duan}, usually with the aim of deriving optimality conditions on the measurements, or as a form amenable to efficient numerical optimization.
Here we are not interested in SDPs as a numerical tool, rather we use the complementary features of the primal and dual programs to propose a solution that turns out to be exact, as we can prove analytically that the upper and lower bounds coincide.

Let us start by deriving the SDP of the problem at hand. We denote by $\ket{0}$ the default state and by $\ket\phi$ the mutated one. Given a sequence of $n$ particles, the change point identification ultimately corresponds to identifying a state within the set of equally likely source states 
$ \{\ket{\Psi_k}\}_{k=1}^n$, where
\begin{equation}
\label{psik}
\ket{\Psi_k}= |\underbrace{0\ldots0}_{k-1}\underbrace{\phi \ldots\phi}_{n-k+1}\rangle
\end{equation}
is associated with the change point occurring at position $k$.
All the discrimination properties of a set of linearly independent pure states are encapsulated in the Gram matrix 
of the set of states~\cite{prl-us,duan}.
For the source states $\ket{\Psi_k}$, it formally reads
%
\begin{equation}
\label{gram-matrix}
G=\sum_{i,j=1}^n\braket{\Psi_i}{\Psi_j}\ketbra{i}{j}=R^\dag R \; ,
\end{equation}
where $\{\ket{i}\}$ is an orthornormal basis of dimension $n$, and \mbox{$R=\sum_k \ketbra{\Psi_k}{k}$}. 
%
%
According to Eq.~\eqref{gram-matrix} one has
$G_{ij}=c^{|i-j|}$, where the overlap $c=\braket{0}{\phi}$ can be assumed to be a positive real number $0\leq c \leq 1$, without loss of generality.

The inverse of $R$ exists due to the linear independence of the set of source states and reads 
$
R^{-1}=\sum_{k=1}^n \ket{k}\!\!\langle \tilde{\Phi}_k |,
$
where $\bra{k}R^{-1}R\ket{l}=\langle \tilde{\Phi}_k\!\ket{\Psi_l}=\delta_{kl}$.
%
%
The tilde reminds us that $|\tilde{\Phi}_k\rangle$ is not a normalized state in general.
Hence we can write
\begin{equation}
\label{omega-inverse}
({R^{-1}})^\dag R^{-1}=\sum_{k=1}^n |\tilde{\Phi}_k\rangle\!\langle\tilde{\Phi}_k|.
\end{equation}
An unambiguous discrimination strategy is characterized by a positive operator valued measure (POVM) consisting of $n$ elements $\{E_{k}\geq 0\}_{k=0}^{n}$ for the hypothesis and an additional element $E_{0}= \openone -\sum_{k=1}^{n}E_{k}\geq 0$ corresponding to the inconclusive outcome. In addition, for the strategy to be unambiguous we need to impose that each outcome $k\in[1,n]$ can only be triggered by one hypothesis $\ket{\Psi_{k}}$,
i.e., $p(k| \Psi_{l})=\tr( E_{k} \ketbrad{\Psi_{l}})=\gamma_{k} \delta_{kl}$. From the observations preceding Eq.~\eqref{omega-inverse} we see that the POVM elements fulfilling this condition are uniquely given by $E_{k}=\gamma_{k} |\tilde{\Phi}_k\rangle\!\langle\tilde{\Phi}_k|$. The so-called \emph{efficiencies}~\cite{duan} $0\leq\gamma_k\leq 1$ are the conditional success probabilities of identifying each source state and are the only free parameters left to optimize the average success probability $P_{s}=\frac{1}{n}\sum_{k=1}^{n}\gamma_{k}$. 
The efficiencies must satisfy the nontrivial constraint given by the completeness relation $E_{0}= \openone -\sum_{k=1}^{n}\gamma_{k}\ketbrad{{\Psi_k}}\geq 0$, which, multiplied by $R^\dag$ (from left) and $R$ (from right), and using Eq.~\eqref{gram-matrix}, results in $G-\Gamma_{\rm D} \geq 0$, with $\Gamma_{\rm D}=\mathrm{diag}\{\gamma_1,\gamma_2,\ldots,\gamma_n\}$. 

We are now in a position to write down the optimization of the overall success probability as an SDP (see Appendix, and also Ref.~\cite{duan}):
\begin{align}
\label{sdp}
P_{\rm s}=\frac{1}{n} & \max_{\Gamma} \tr \Gamma 	\hspace{.2cm} \nonumber \\
\mbox{ subject to} \hspace{.2cm} & G-\Gamma_{\rm D} \geq 0 \;, \\
& \Gamma \geq 0 \; .\nonumber
\end{align} 
%
For convenience, we have relaxed the SDP to operate over a matrix variable $\Gamma$, whose diagonal is $\Gamma_D$.
This SDP formally coincides with the type of
classical optimization problems known as trace factor analysis~\cite{trace-analysis}, where, e.g. 
one wants to bound the amount of noise compatible with an observed correlation matrix.
Note that the  lowest  eigenvalue of $G$ directly yields a lower bound for $P_{\rm s}$,
because the two inequalities in Eq.~\eqref{sdp} are
trivially satisfied. 
In symmetric discrimination scenarios, e.g., when the hypotheses are generated by the action of a unitary operation $U$ satisfying $U^n = \openone$, all 
the efficiencies coincide, and the lowest eigenvalue
is in fact the exact success probability.  
Instead, the change point problem has a privileged direction from left to right, 
hence one expects a more involved solution.

%
The primal SDP, Eq.~\eqref{sdp}, entails an optimization that is hard to tackle analytically. In contrast, its dual form is much more eloquent (see Appendix):
\begin{align}
\label{dual}
P_{\rm s}=\frac{1}{n} & \min_{Z}  \; \tr G Z 	\hspace{.2cm}  \nonumber \\
\mbox{ subject to} \hspace{.2cm} & Z_{kk}\geq 1 \ ,\quad k= 1,\ldots,n \;,\nonumber \\
& Z\geq 0 \; .
\end{align}
Its structure facilitates the means to introduce an analytical ansatz of the dual variable $Z$.
Notice that any choice of a positive semidefinite operator $Z$ with diagonal terms larger or equal than unity yields an upper bound to the 
success
probability. The minimization of the objective function in Eq.~\eqref{dual} suggests taking the most straightforward choice,  a rank-one projector $Z=\ketbrad{u}$, and to consider the components of $\ket{u}$  to have the minimal value with alternating signs,  $u_k=\braket{k}{u}=(-1)^{k+1}, \ k=1,\ldots,n$. 
Remarkably, as we prove below, this ansatz attains the optimal success probability for values of the overlap up to a critical threshold $c^*$, that is, in the interval $0\leq c\leq c^*$.
The existence of such a threshold is related to the convex structure of the problem: up to $c^*$ the optimal probability is attained by a boundary point of the feasibility region of Eq.~\eqref{dual} (i.e., it fulfills $Z_{kk}=1$), 
but for $c > c^*$ the optimal $Z$ demands some of its diagonal components to be strictly larger than the unity; thus, it becomes an interior point.
This notwithstanding, a slight modification of the vector $\ket{u}$ leads to the optimal success probability for $c>c^*$,
hence completing the solution for any value of the overlap.

Let us start by analyzing  the first regime, $c<c^*$. 
The ansatz with $u_k=(-1)^{k+1}$  provides, in principle, an upper bound to the success probability:
\begin{eqnarray}
\label{ps-upper}
P_{\rm s}&\leq& \frac{1}{n}\bra{u}G\ket{u} =
                                      \frac{1}{n}\sum_{i,j=1}^n (-c)^{|i-j|} \nonumber \\
                                     & =&\frac{1-c}{1+c}+ \frac{1}{n} \frac{2c\left[1-(-c)^n\right]}{(1+c)^2} \equiv P_{\rm s}^{\rm I}\,.
\end{eqnarray}
In order to prove the tightness of this upper bound, it is enough to find a feasible point of the primal problem in  Eq. \eqref{sdp} that attains the same value $P_{\rm s}^{\rm I}$.
For this purpose,  given ansatz $Z=\ketbrad{u}$ of the dual problem, we construct an ansatz of the primal problem $\Gamma={\rm diag}\{\gamma_1,\gamma_2,\ldots,\gamma_n\}$ with
\begin{align}
\label{efficiencies}
\gamma_k = u_k \sum_{j=1}^n G_{kj} u_j=\sum_{j=1}^n (-c)^{|k-j|} \,,\quad k=1,\ldots,n \,.
\end{align}
These  \emph{induced efficiencies}, when plugged into the objective function of Eq.~\eqref{sdp}, trivially give $P_{\rm s}^{\rm I}$. Hence, it just remains to prove that $\Gamma$
is a feasible solution,
i.e., that it satisfies the conditions $\Gamma\geq 0$ and  \mbox{$G-\Gamma_{\rm D}\geq 0$}. 
%
\begin{figure}
   \centering
   \includegraphics[width=3.4 in]{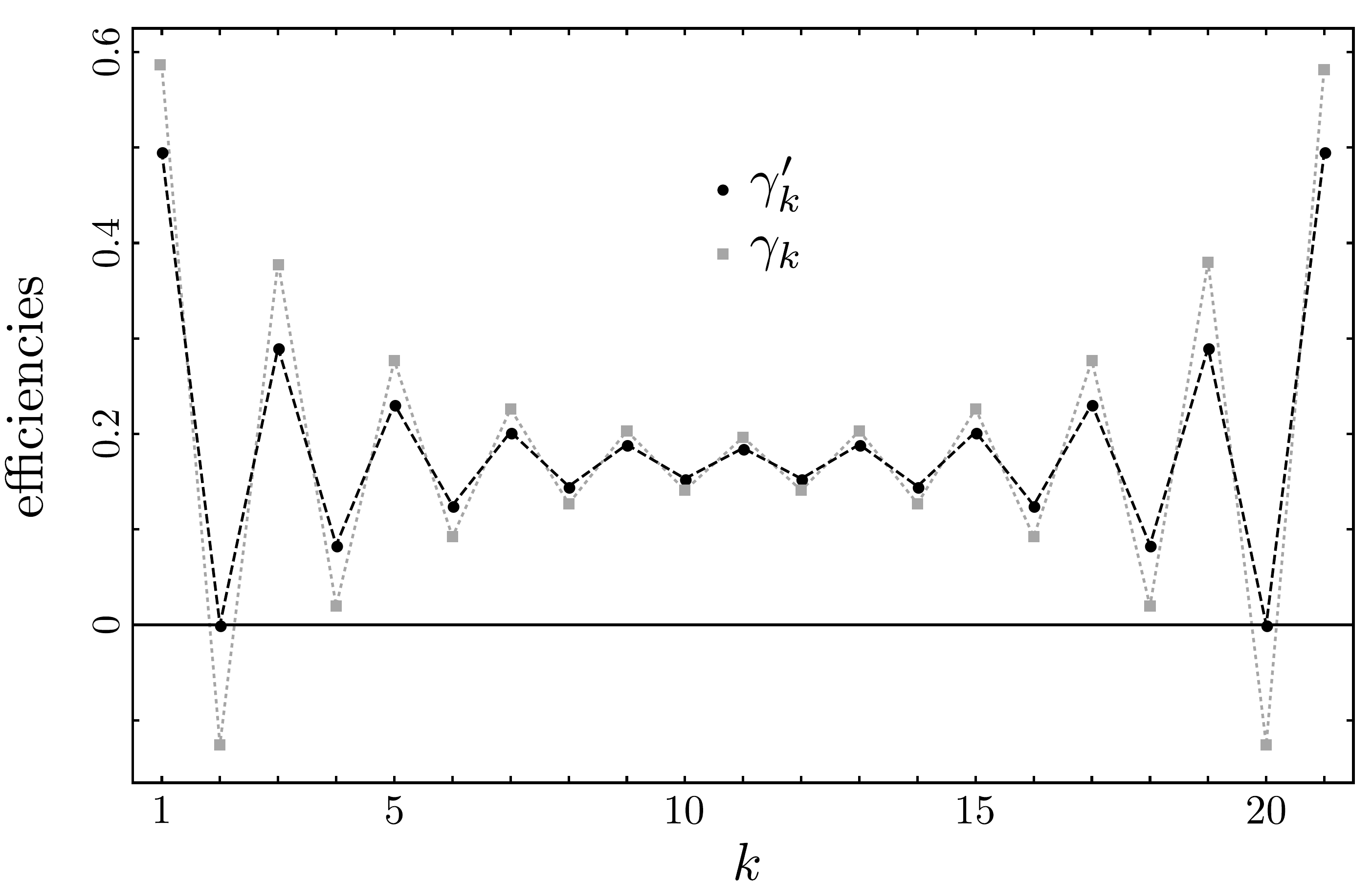} 
\caption {
Conditional success probabilities $\gamma'_k$ (black) for $n=20$ and $c=0.7>c^* \simeq 0.62$.
The efficiencies $\gamma'_2$ and $\gamma'_ {19}$ vanish and the rest are all positive, as required in Eq.~\eqref{sdp}. 
The oscillations in the values of  the efficiencies attenuate as one approaches the central values of $k$. We also depict (gray)
the efficiencies $\gamma_k$ as would be given by Eq.~\eqref{efficiencies}. The meaningful ones in the region $c\geq c^*$, given by Eq.~\eqref{efficiencies-p}, 
can be seen as a compression of the oscillations exhibited by the unphysical efficiencies $\gamma_k$.}
\label{fig:1}
\end{figure}

We will shortly see that this is indeed the case for overlaps smaller than a critical value.
As can be seen in Fig.~\ref{fig:1}, the value of the induced efficiencies $\gamma_k$ oscillate with the position $k$ of the change point.
This is a rather unexpected behavior. The optimal protocol favors the identification of some hypotheses at the expense of 
penalizing others in an alternating way. 
The end hypotheses, corresponding to $k=1$ and $k=n$, have the highest efficiencies, while 
$k=2$ and $k=n-1$ have the lowest. 
To demonstrate that $\Gamma_{\rm D}\geq 0$
it is enough to prove that indeed $\gamma_k\geq \gamma_2=\gamma_{n-1}, \ \ \forall k\neq 2,n-1$, and 
that $\gamma_2=\gamma_{n-1}\geq 0$~(see Appendix). 
The efficiencies for the central hypotheses, i.e., around
$k\sim n/2$, are essentially constant for large values of $n$. The oscillations become noticeable at the end points and can be seen as a border effect. It is interesting to note that,
if we consider the symmetric change point with periodic boundary conditions, these oscillations disappear~\cite{preparation}.
At the critical overlap (and beyond as we show below), $\gamma_2$ and $\gamma_{n-1}$ vanish, i.e., it pays off to give up the identification of these two points.
This behavior, although surprising,  is not totally unusual: in the unambiguous discrimination of two quantum states, if the prior probability of one of them is above some threshold, it is preferable to forget about the other state and design a POVM that either confirms the state with higher probability or gives an uninformative outcome. 
Then, as argued, the equation $\gamma_2=0$ (or, equivalently, $\gamma_{n-1} = 0$) determines the critical overlap $c^*$. Using 
Eq.~\eqref{efficiencies}, we have 
\begin{equation}
 \label{gamma2}
 \gamma_2=\gamma_{n-1}=\frac{1-c-c^2-(-c)^{n-1}}{1+c}
 \end{equation} 
and the equation for $c^*$ can be written as%
\begin{equation}
\label{c-critical}
1-c^*-{c^*}^2-(-c^*)^{n-1}=0\,.
\end{equation}
%
For large $n$ the critical overlap $c^*$  is given by the inverse of the golden ratio, and actually, since the correction is exponentially small, $n$ need not be very large to achieve this value.

As for the remaining condition $A:= G-\Gamma_{\rm D} \geq 0$, 
note that $A$ 
has at least one zero 
eigenvalue, for  $\bra{u}G\ket{u}=\bra{u} \Gamma_{\rm D}\ket{u}$. Then, 
if the first $n-1$ leading principal minors  
\footnote{The $k$th leading principal minor of a matrix $A$ is the determinant of the submatrix $[A]_{ij}$, with $i=1,\ldots,k$ and $ j=1,\ldots, k.$} 
 of $A$ 
are positive and $\det A=0$, 
$A$ is positive semidefinite \cite{horn}. 
One can prove that, up to the critical overlap, this is so (see details in Appendix).
%
Hence, in the interval $0\leq c\leq c^*$, $\Gamma_{\rm D}$ is a feasible solution of the SDP~\eqref{sdp} and thus $P_{\rm s}^{\rm I}$ is the optimal success probability.

For $c\geq c^*$, the ansatz for $\ket{u}$ has to be modified to keep all the efficiencies positive. We do the minimal modification and 
consider (in vector notation) $\vec{u}'=(1,-b,1,-1,\ldots, (-1)^{n-1} b, (-1)^n)$, and define the new induced efficiencies  $\gamma'_k$ following the same prescription as in Eq. \eqref{efficiencies}. These efficiencies can be shown to have the same behavior as $\gamma_{k}$, in particular they fulfil $\gamma'_k\geq \gamma'_2=\gamma'_{n-1}$~(see Appendix).
Hence, in order to warrant the positivity condition on $\Gamma'$, the parameter $b$ is chosen is such a way that nullifies $\gamma'_2$ and 
$\gamma'_{n-1}$,
\begin{equation}
b=1-\frac{\gamma_2}{1+(-c)^{n-3}}\,.
\end{equation}
The modified efficiencies then read
\begin{equation}
\label{efficiencies-p}
\gamma'_k=\gamma_k-(1-b)\left[(-c)^{|k-2|}+(-c)^{|n-k-1|}\right].
\end{equation}
Figure~\ref{fig:1} illustrates how all the efficiencies $\gamma'_k$ with \mbox{$k\neq 2,n-1$} are positive while $\gamma'_2$ and $\gamma'_{n-1}$  remain null, for all
$c\geq c^*$.  The success probability can be readily computed from Eq.~\eqref{efficiencies-p} to give
\begin{equation}
\label{ps-p}
P_{\rm s}^{\rm II}=P_{\rm s}^{\rm I}+\Delta \,, \mbox{ with } \Delta=-\frac{2}{n} \frac{\gamma_2^2}{1+(-c)^{n-3}}\,.
\end{equation}
%
For large $n$ one has
\begin{equation}
\Delta \simeq-\frac{2}{n} \left(\frac{1-c-c^2}{1+c}\right)^2 \,.
\end{equation}

Again, we still have to verify that $\Gamma'={\rm diag}\{\gamma'_1,\gamma'_2,\ldots,\gamma'_n\}$ yields a semidefinite positive operator $A'=G-\Gamma'\geq 0$. 
It is easy to check numerically that $A'\geq 0$ for any value
of $c$ in the region $c^*<c\leq 1$, but 
the analytical proof is much more involved, as now the first and second rows and columns of $A'$ are linearly dependent and, hence, all its leading principal minors vanish. Then, in principle, one should prove the positivity of \textit{all} principal
minors~\cite{horn}, and there are $2^n-1$ of them instead of the $n$ leading minors. This may seem a prohibitive task, but the following observation greatly simplifies  the problem. It is easy to check that the 
kernel of $A'$ is 
spanned by the vectors $\vec{v}_1=(1,-c,0,\ldots,0)$, $\vec{v}_{n-2}=(0,\ldots,0,-c,1)$, and the modified alternating vector $\vec{u}'$.
Taking the intermediate basis vectors $\vec{v}_2=(0,0,1,0,\ldots,0)$, $\vec{v}_3=(0,0,0,1,0,\ldots,0),\ldots,  \vec{v}_{n-3}=(0,\ldots,0,1,0,0)$, and defining $\vec{v}_1^\bot:=(c,1,0,\ldots,0)$ and $\vec{v}_{n-2}^\bot:=(0,\ldots,0,1,c)$, one can construct the operator $P$ that has the vectors $\vec{v}_1^\bot,\vec{v}_2,\ldots,\vec{v}_{n-3},\vec{v}_{n-2}^\bot$ as rows and 
removes the part of the kernel of $A'$ corresponding to vectors $\vec{v}_1$ and $\vec{v}_{n-2}$.
The resulting operator, $B:=P A' P^{T}$,
has dimensions $(n-2)\times(n-2)$,
the same number of positive eigenvalues as $A'$
(the eigenvalues are not exactly the same because for simplicity we have not normalized $\vec{v}_1$ and $\vec{v}_{n-2}$), and one zero eigenvalue corresponding to $\vec{u}'$.
Since $B\geq 0\Rightarrow A'\geq 0$, at this point we can use the easier criterion for positive semidefiniteness of $B$ that concerns only its leading principal minors, in the same fashion as we proved that $A\geq 0$ for $0\leq c\leq c^*$. 
The interested reader can find the the technical details and the complete  proof in the Appendix.
 
 \begin{figure}
   \centering
   \includegraphics[width=3.4 in]{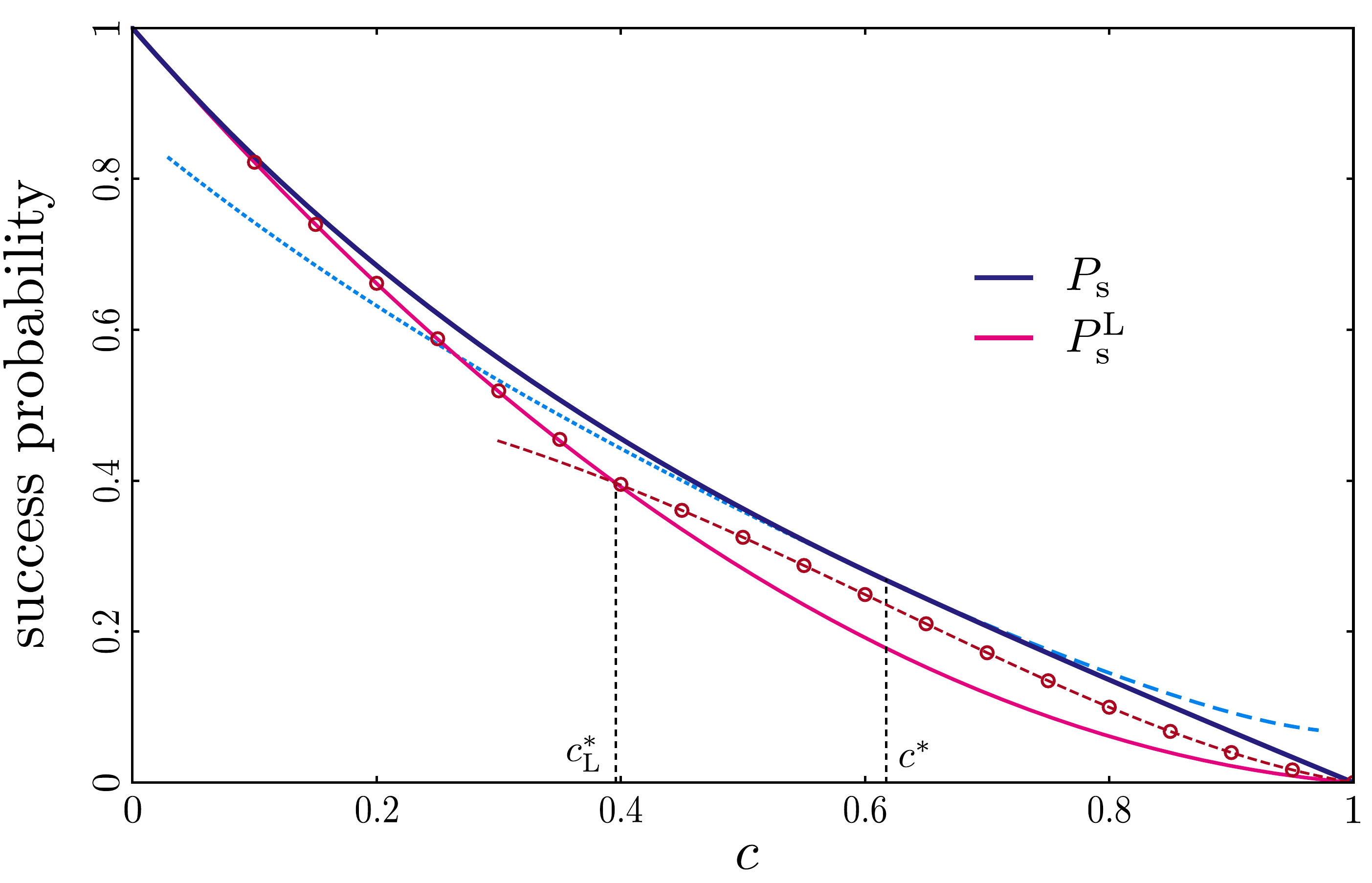} 
\caption {Probability of exact identification of the change point as a function of  $c=|\braket{0}\phi|$ for $n=15$. The solid (dark blue) 
line is the 
exact piecewise function of the success probability for $0\leq c\leq 1$. Notice that $P_{\rm s}$, although piecewise defined, is a differentiable function.
The dashed (light blue) curve  is the continuation of  the success probability 
$P_{\rm s}^{\rm I}$, Eq.~\eqref{ps-upper}, into the region $c\geq c^*$ where it is not the valid solution. The dotted (light blue) curve is the continuation of 
$P_{\rm s}^{\rm II}$, Eq.~\eqref{ps-p}, into the region $0\leq c\leq c^*$ where it is also not an acceptable solution. The solid 
(pink) curve is the success probability of the local protocol, $P_{\rm s}^{\rm L}$, given by Eq.~\eqref{ps-local}. The red circles are the results of a numerical optimization over a more general class of local protocols, where we considered the local efficiencies as independent parameters. Below $c_{\rm L}^*$ the numerical points essentially coincide with $P_{\rm s}^{\rm L}$, which corresponds to a strategy of equal local efficiencies (the difference is of the order $10^{-3}$). Above this threshold, they match the success probability of a strategy that exclusively detects either $\ket{0}$ or $\ket{\phi}$ at each position, in an alternate way (dashed red). The optimal local strategy sharply transitions between these two regimes at $c_{\rm L}^*$.
 }
\label{fig:2}
\end{figure} 

The analytical expression of the optimal success probability is a piecewise function given by Eq.~\eqref{ps-upper} for $0\leq c\le c^*$ and
Eq.~\eqref{ps-p} for $c^*< c \leq 1$. This function, depicted in Fig.~\ref{fig:2}, is differentiable, but has a discontinuity in the 
second derivative at the critical value $c^*$ (see Ref.~\cite{bergou} where a similar effect is observed).
Note  that, for $n\to \infty$, the success probability converges to a constant $P_{\rm s}\simeq (1-c)/(1+c)$ (valid for all $c$), which is the lowest eigenvalue of the Gram matrix in this asymptotic regime~\cite{prl-us}. 
%

For $c<1$ the factors $c^n$ vanish exponentially fast with $n$. Disregarding these factors the success probability is
\begin{equation}
P_{\rm s}\simeq
\begin{cases}
\frac{1-c}{1+c}+ \frac{1}{n} \frac{2c}{(1+c)^2} \ \  \ \ \ \ \ \  0\leq c \leq c^*= (\sqrt{5}-1)/2\\
\frac{1-c}{1+c}+ \frac{1}{n} \frac{2c}{(1+c)^2}-\frac{2}{n} \left(\frac{1-c-c^2}{1+c}\right)^2 \ \ \ \  c^* < c\leq 1\,.
\end{cases}
\end{equation}

It is interesting to compare the optimal success probability, attained by a collective measurement on the whole sequence of particles, with that obtained with local protocols, i.e., those where each particle is measured individually.  
Online strategies are particularly interesting among this class of protocols: in these, an observer measures the particles sequentially and, with some probability, detects the change point as soon as it occurs. A simple online strategy that one can consider consists in performing locally optimal unambiguous discrimination measurements for the states $\ket{0}$ and $\ket{\phi}$ with equal priors. Such strategy enforces equal efficiencies, $\gamma_{\rm L}=1-c$, for all possible change points. This is a reasonable assumption, specially in the limit of large $n$ where there are no boundary effects and where, for the optimal collective strategy, efficiencies approach a constant value (except for a few points near the boundary).
%
%
The probability of a correct identification is the probability of obtaining two conclusive outcomes just before and at the change point, 
hence, $P^{\rm L}_{\rm s}\approx (1-c)^2$ . For sequences of a given length $n$, if we take into account that for the first and last change point we only need one inconclusive outcome, we have the refined expression
\begin{align}
\label{ps-local}
P^{\rm L}_{\rm s} & =\frac{n-2}{n} \gamma_{\rm L}^2+\frac{2}{n}\gamma_{\rm L} =\frac{n-2}{n}(1-c)^2+\frac{2(1-c)}{n}\nonumber \\
             &=(1-c)^2+\frac{2c(1-c)}{n}\,.
\end{align}
As expected, the ratio $(P_{\rm s}-P^{\rm L}_{\rm s})/P_{\rm s}\simeq c^2$ is positive. Notice 
that for small values of $c$, i.e., very orthogonal 
states,  $P^{\rm L}_{\rm s}\simeq P_{\rm s}$, which tells us that the full quantum correlations of the global measurements do not provide too much advantage over this simple 
online strategy in this regime.

One can, of course, devise more sophisticated local strategies. A reasonable approach is to optimize over the local efficiencies to maximize the overall success probability. Note that, as opposed to the strategy described above, this one, in principle, will 
have the 
specific length of the sequence of particles
embedded in its design. We carry out this optimization numerically and observe that, up to a certain overlap, $c^*_{\rm L}$, the found efficiencies are indeed essentially constant for the majority of possible change points. However, beyond this overlap, contrary to 
our naive ansatz of equal efficiencies, the oscillatory behavior emerges again: the local unambiguous measurements become two-outcome measurements and are completely biased to detect only one of the states, $\ket{0}$ or $\ket{\phi}$, in an alternate fashion. 
This results in an improvement over $P_{\rm s}^{\rm L}$ for $c>c_{\rm L}^*$, although not large enough to reach the performance of the collective strategy.
The transition between these two regimes is sharp.
We illustrate this phenomenon in Fig.~\ref{fig:2} for $n=15$. In the Appendix we discuss in more detail this strategy and give a semi-analytical proof that for large $n$ the local critical overlap that determines the change of regime is $c^*_{\rm L}\approx \sqrt{2}-1$.


In summary, we have computed the optimal probability of exact identification of a quantum change point for any length of the sequence of particles and for 
any value of the overlap between the default and mutated states.  The SDP formalism provided the necessary insight to find the analytical solution.
The exact identification of change points thus constitutes one of the unique examples of multihypothesis discrimination problems where a closed solution can be found beyond symmetric cases.
The optimal protocol exhibits an unexpected and nontrivial oscillatory behavior of the efficiencies as a function 
of the position of the change point, illustrated in Fig.~\ref{fig:1}. 
More general scenarios with, e.g., several change points seem to be addressable with the results presented here and are currently under investigation.
We have also described a simple online strategy capable of unambiguously detecting the change point, and have shown that the optimal protocol substantially outperforms it, especially for states with a rather large overlap close to the critical threshold $c^*$.
Exploring more sophisticated local protocols, we have seen that the performance of the optimal protocol remains unchallenged.

This research was supported by the Spanish MINECO
through contracts FIS2013-40627-P, FIS2015-67161-P \& FIS2016-80681-P, the ERC Starting Grant 258647/GEDENTQOPT and Consolidator Grant 683107/TempoQ, the DFG, and the Generalitat
de Catalunya CIRIT contract 2014-SGR966. We thank Emilio Bagan and Janos Bergou for useful discussions.

\bibliographystyle{unsrt}

\appendix
\section{Primal and dual SDP}
A very convenient way of writing a SDP for quantum mechanical problems is \cite{watrous_supp}
\begin{align}
\label{sdp-primal-1}
\max & \,\tr A X \nonumber \\
&\Phi[X]=B \\
&X\geq 0 \,,\nonumber
\end{align}
where $X$ is the unknown matrix and $A$, $B$ and $\Phi[\bullet]$ are determined by the problem to be solved. Here $\Phi[\bullet]$ is a linear hermiticity preserving map and $A$, $B$ and $X$ are hermitian matrices.  
Notice that this formulation differs from the standard canonical form of \cite{vandenberghe_supp},
although both are completely equivalent. 

The dual version of Eq.~\eqref{sdp-primal-1} is~\cite{watrous_supp}
\begin{align}
\label{sdp-dual-1}
\min & \, \tr B Y \nonumber \\
&\Phi^\dag[Y]\geq A\,.
\end{align}
Now the variable is $Y$, and the dual map $\Phi^\dag[\bullet]$ is defined from the condition $\tr Y \Phi[X] =\tr \Phi^\dag[Y] X$.

If instead of an equality constraint one has an inequality, 
as is the case in unambiguous discrimination, 
\begin{align}
\label{sdp-primal-2}
\max & \,\tr A X \nonumber \\
&\Phi[X]\leq B \\
&X\geq 0 \,,\nonumber
\end{align}
 one simply introduces a slack variable $Z\geq 0$, writes $\tilde{X}=X\bigoplus Z$, defines the map $\tilde{\Phi}[\tilde{X}]=\Phi[X]+Z$ and makes the extension $\tilde{A}=A\bigoplus \mathbf{0}$, where $\mathbf{0}$
is the null matrix. These definitions transform the problem into the standard form  for $\tilde{A}, \tilde{\Phi}[\bullet]$ and $B$ [cf. Eq.~\eqref{sdp-primal-1}]:
\begin{align}
\label{sdp-primal-3}
\max & \,\tr \tilde{A}\tilde{X}=\tr A X  \nonumber \\
&\tilde{\Phi}[\tilde{X}]=\Phi[X]+Z=B \\
&\tilde{X}\geq 0  \Leftrightarrow X,Z\geq 0\,.\nonumber
\end{align}
The dual version can be directly read from Eq.~\eqref{sdp-dual-1}:
\begin{align}
\label{sdp-dual-2}
\min & \, \tr B Y \nonumber \\
&\Phi^\dag[Y]\geq A  \\
&Y\geq 0 \,.\nonumber
\end{align}
Notice the beautiful duality between Eqs.~\eqref{sdp-primal-2} and \eqref{sdp-dual-2}.

In the unambiguous problem of the main text, Eq.~\eqref{sdp},  $A=\openone$, $B=G$, $X=\Gamma$, $\Phi[X]=X_{\rm D}$, where
$X_{\rm D}=\mathrm{diag}\{X_{11},X_{22},\ldots,X_{nn}\}$, and the dual map  just reads
 $\Phi^\dag[Y]=Y_{\rm D}$.

 \section{Proofs of optimality}

In this section we prove the optimality of our solution, that is, we prove that Eqs.~\eqref{ps-upper} and \eqref{ps-p} correspond to the exact optimal unambiguous discrimination probability for arbitrary $n$ for $0\leq c \leq c^*$ and $c^* < c \leq 1$, respectively. We begin with the first region. As argued in the main text,
an appealing ansatz for $Z$ of the dual  SDP problem, Eq.~\eqref{dual}, is $Z=\ketbrad{u}$, with $\ket{u}$ being a vector with components $u_k=(-1)^{k+1}$.  The induced efficiencies $\gamma_k$ then read
\begin{equation}
\label{efficiencies-sup-1}
\gamma_k =\sum_{j=1}^n (-c)^{|j-k|}\,.
\end{equation}
We now check whether these efficiencies satisfy the SDP constraint $\Gamma_{\rm D}\geq 0$. Notice that $\gamma_k=\gamma_{n-k+1}$, hence one needs to consider only $k=1,\ldots,\lceil n/2\rceil $ (from now on and to ease the presentation, we omit references to the identical symmetric efficiencies). It is easy to prove that $\gamma_{2}=\gamma_{n-1}<\gamma_k$ for all $k\neq 2,n-1$. Then, if
one has $\gamma_{2}>0$, the positivity 
condition $\Gamma_{\rm D}\geq 0$ is automatically satisfied.
We observe  that 
\begin{equation}
\gamma_k-\gamma_2=\sum_{j=2}^{k-1} (-1)^j[c^j-c^{n-j}]=\sum_{j=2}^{k-1}(-1)^j a_j\,.
\end{equation}
Since $a_j \geq 0$ and $a_j>a_{j+1}$ for $k\leq \lceil n/2\rceil$, one has that  $\gamma_k-\gamma_2 \geq 0$.
Also from $\gamma_2=[1-c-c^2-(-c)^n]/(1+c)$ [cf. Eq.~\eqref{gamma2}] one has 
$\gamma_2 \leq 0$ for $c^*\leq c \leq 1$, where the equality is attained at $c=c^*$, i.e., at the positive root of the equation
\begin{equation}
\label{c-cut}
1-c-c^2-(-c)^{n-1}=0\,.
\end{equation}
Thus, in the region $0\leq c\leq c^*$, the efficiencies given by Eq.~\eqref{efficiencies-sup-1} are all positive. Interestingly, in the limit $n\to\infty$ we can neglect the exponential term in Eq.~\eqref{c-cut}, and $c^*$ becomes the inverse of the golden ratio, that is, $c^*\to (\sqrt{5}-1)/2$.

Next we have to demonstrate  the positivity condition $A=G-\Gamma\geq 0$.  Recall that a matrix $A$  is positive semi-definite if the
 first $n -1$ leading principal minors  of A are positive and $\det A\geq 0$ \cite{horn_supp}. Denoting by $M_k$ the leading minor
of order $k$ of the matrix 
$A=G-\Gamma$,  one can easily check that
\begin{align}
\label{minors-1}
\eta_k:=&\frac{M_{k+1}}{M_k} \nonumber \\
=&\frac{c+(-c)^{n-k}}{(1+c)[1-(-c)^{n-k}]}[1-c-(-c)^{k+1}-(-c)^{n-k}].
\end{align}
Defining $M_0=1$, Eq.~\eqref{minors-1}  holds for any $k$ and $M_k=\prod_{s=0}^{k-1} \eta_s$. Therefore one just has to prove that $\eta_k>0$ for all $k\in [1,n-2]$ and that $\eta_{n-1}\geq 0$. Actually, it is trivially seen from Eq.~\eqref{minors-1} that $\eta_{n-1}=0$, hence  $\det A=0$. Notice that the fraction factor in Eq.~\eqref{minors-1} is  positive for $k\in[1, n-2]$, so  only the last factor
\begin{equation}
1-c-(-c)^{k+1}-(-c)^{n-k}=1-c-c^2-(-c)^{n-1}+ \Delta_k\,,
\end{equation}
where 
\begin{equation}
\label{Deltak}
\Delta_k=\left[c^2+(-1)^k c^{k+1}\right]+ (-1)^{n-1}\left[c^{n-1}+(-1)^k c^{n-k}\right],
\end{equation}
is relevant for the positivity of $\eta_k$.
Recall that $1-c-c^2-(-c)^{n-1}$ is positive for  $0\leq c<c^*$ [see Eq.~\eqref{c-cut}].
Finally notice that $\Delta_k=\Delta_{n-k+1}$ and that the first term in Eq.~\eqref{Deltak} is always non-negative and bigger or equal than the absolute value of the second term for $k\leq \lceil n/2 \rceil$. When $k$ is even this is clear, and when 
$k$ is odd just notice that $c^s-c^m=(1-c)\sum_{j=s}^{m-1} c^j$.
Thus we  have  $\Delta_k\geq 0$,  $k\in[0,n-1]$, which finishes the proof for $0\leq c\leq c^*$.

We now complete the proof by focusing on the overlap interval $c^* < c \leq 1$. We show that the efficiencies in Eq.~\eqref{efficiencies-p} yield a feasible solution of the SDP problem~\eqref{sdp} and, therefore, $P_{\rm s}^{\rm II}$ is the optimal success probability of identification. As outlined in the main text, we need to prove the positivity of the operator $A'=G-\Gamma'$. The operator $A'$ has three zero eigenvalues, with an associated eigenspace spanned by the vectors
\begin{align}
&\vec{v}_1 =(1,-c,0,\ldots,0) \,,\\
&\vec{v}_{n-2} =(0,\ldots,0,-c,1) \,,\\
&\vec{u}' =(1,-b,1,-1,\ldots,(-1)^{n-1}b,(-1)^n)\,,
\end{align}
where
\begin{equation}
b=c \left( 1+\frac{1+(-c)^{n-5}c}{(1+c) [1+(-c)^{n-3}]} \right)\,.
\end{equation}
Let $P$ be an operator that removes the zero-eigenvalue subspace ${\rm span}\{\vec{v}_1,\vec{v}_{n-2}\}$ from $A'$, and $B:=P A' P^\dagger$ be the result after the action of $P$. Note that $B$ still contains one zero eigenvalue, corresponding to the vector $\vec{u}'$. If $B \geq 0$, then $A'\geq 0$. One can check numerically that all leading principal minors of $B$, denoted by $M'_k$, where $k\in[1,n-2]$, are positive except for $M'_{n-2}$. This makes the problem much more tractable than trying to prove directly that $A'$ is positive semidefinite, since, according to Silvester's criterion~\cite{horn_supp}, it is enough to show that the leading principal minors (but $M'_{n-2}$) are positive.

It is tedious but straightforward to deduce the explicit form of $M'_k$ by induction based on examples for small values of $n$. It reads $M'_k = R_k \cdot S_k$, where
\begin{align}
R_k := &\frac{(1+c^2)^2}{c^{\frac{k(k-1)}{2}}[c^3-(-c)^n]^{k-1}} (1-c)^{\lfloor \frac{k-1}{2} \rfloor +k-1} \nonumber\\
&\times \prod_{s=3}^k [c^{s+2}-(-1)^s(-c)^n] \nonumber\\
&\times \{c^{k+3}+(-1)^k(-c)^n [1-c-(-c)^k]\} \,,\label{Rk}\\
S_k := &\left[\prod_{m=0}^{\lfloor\frac{k-2}{2}\rfloor} \sum_{j=0}^{2m}(-c)^{j}\right]
\left(	\prod_{r=0}^{\lfloor\frac{k-3}{2}\rfloor} \sum_{i=0}^{r} c^{2i}	\right) \,. \label{Sk}
\end{align}
We want to show that $M'_k > 0$, $k\in[1,n-3]$ for $c^* < c \leq 1$. To this end, we can get rid of the trivially positive factors in $M'_k$, namely the first line of Eq.~\eqref{Rk} and the second factor in Eq.~\eqref{Sk}. Let us call the remaining terms $R_k^2$, $R_k^3$ and $S_k^1$, where the superindex marks the appearance order in Eqs.~\eqref{Rk} and \eqref{Sk}. The term $S_k^1$ is a product of positive sums, each of them being strictly smaller than the previous one as the index $m$ increases. The smallest possible sum that can be added to the product is then $\sum_{j=0}^{\infty} (-c)^j = 1/(1+c)$, which is also positive. The positivity of $R_k^2$ and $R_k^3$ becomes apparent by taking into account that $k\in[1,n-3]$. Since $M'_k > 0$ and $M'_{n-2} = \det{B} = 0$, we conclude that $B\geq 0$ and so is $A'$.

%

\section{Local strategies}

The conditional probability of unambiguous discrimination of state $\ket{0}$  with prior probability $\eta_1$ is $1- c\sqrt{\eta_2/\eta_1}$ 
and $1- c\sqrt{\eta_1/\eta_2}$ for state $\ket{\phi}$~\cite{barnett_supp}. Hence, the parameter characterizing the unambiguous measurement of system $k$ is  the weight
$x_k:=\sqrt{\eta^{(k)}_2/\eta^{(k)}_1}$. The probability of exact identification at position $k$ is $(1-c x_{k-1})(1-c/x_k)$,  
where the constraint $c\leq x_k \leq 1/c$ has to be taken into account. 
The average success probability reads
%
\begin{equation}
\label{pl}
P^{\rm L}_{\rm s}(\mathbf{x})=\frac{1}{n}\sum_{k=0}^{n-1} (1-c x_k)\left(1-\frac{c}{x_{k+1}}\right), 
\end{equation}
where $\mathbf{x}=\{x_0,x_1,x_2,\ldots x_n\}$, and the boundary conditions have been taken into account by setting $x_0=0$ (i.e., the state before the first position  
 is $\ket{0}$ for sure) and $1/x_{n}=0$ (i.e., we are sure that  the 
last particle is in state $\ket{\phi}$). For overlaps below a critical value $c_{\rm L}^*$, a numerical maximization of Eq.~\eqref{pl}
over $\mathbf{x}$ shows that $x_k\approx 1$ for all $k$ away from $k=1$ and $k=n-1$. 
However, for $c\geq c^*_{\rm L}$ this solution ceases to be an overall maximum. The choice of the extremal values 
$x_1=1/c,\,x_2=c,\,x_3=1/c\ldots$ yields a higher success probability. This  strategy corresponds to a two-outcome measurement that detects unambiguously
only one of the states, starting with $\ket{\phi}$ and followed by $\ket{0}$ in an alternated way. Hence, roughly only half of the change points are detected. The transition from one strategy to the other is sharp, and the crossing of the curves of $P^{\rm L}_{\rm s}(\mathbf{x})$ for the two strategies determines the value $c^*_{\rm L}$. 
Using Eq.~\eqref{pl}, in the large $n$ limit we obtain
\begin{equation}
\label{cl}
(1-c)^2=\frac{(1-c^2)^2}{2} \to c^*_{\rm L}\approx\sqrt{2}-1\,.
\end{equation}
There is a minor subtlety when the number of particles is even. The values of the weights start with $x_1=1/c$, i.e., with a measurement that detects $\ket{\phi}$, and have to end with $x_{n-1}=c$, i.e., a measurement detecting only $\ket{0}$. When $n$ is even  these boundaries cannot be matched unless some weight is repeated or takes another value.  The optimal solution in that case is the alternated value of the extreme weights  but one, with 
$x_k=1$, at one odd position $k=2j+1$, e.g., $x_1=1/c$, $x_2=c$, $\ldots,$ $x_{2j}=c$,  $x_{2j+1}=1$, $x_{2j+2}=1/c$, $\ldots,$ $x_{n-1}=c$. The solution  is degenerate in the sense that the value $x_k=1$, at any odd position except $k=1$ and $k=n-1$, yields the same optimal success probability. Of course, this correction is not important to compute the success probability and the critical 
overlap value for large $n$, as it is enough to consider strings with odd number of particles, which do not present this particularity.

One could refine further the local strategy and consider Bayesian updating protocols. However,
unlike in the minimum error case \cite{prl-us-p_supp}, where preceding outcomes can be quite informative, here all the inconclusive outcomes are of little use, and only the last conclusive outcome is expected to provide some useful information. 
This  matter is currently under investigation.

 \bibliographystyle{unsrt}

\end{document}